\documentclass[twocolumn,aps,prl,superscriptaddress,longbibliography]{revtex4}
\usepackage{color}
\usepackage{amsmath}
\usepackage{amssymb}
\usepackage{graphicx}
\usepackage{esint}
\usepackage[unicode=true,bookmarks=false,breaklinks=true,pdfborder={0 0 0},backref=false,colorlinks=true]{hyperref}

\usepackage{graphicx}
\usepackage{theorem}
\usepackage{dcolumn}
\usepackage{amssymb}
\usepackage{amsmath}
\usepackage{amsfonts}
\usepackage{bm}
\usepackage{tikz}
\usepackage{pdfpages}
\usepackage{verbatim} 
\usepackage{soul}

\begin{document}

\title{Experimental discovery of nodal chains}
\author{Qinghui Yan$^\dagger$}
\affiliation{Institute of Physics, Chinese Academy of Sciences/Beijing National Laboratory for Condensed Matter Physics, Beijing 100190, China}
\affiliation{State Key Laboratory of Modern Optical Instrumentation, College of Information Science and Electronic Engineering, Zhejiang University, Hangzhou 310027, China}
\author{Rongjuan Liu$^\dagger$}
\affiliation{Institute of Physics, Chinese Academy of Sciences/Beijing National Laboratory for Condensed Matter Physics, Beijing 100190, China}
\author{Zhongbo Yan}
\affiliation{Institute for Advanced Study, Tsinghua University, Beijing 100084, China}
\author{Boyuan Liu}
\affiliation{Institute of Physics, Chinese Academy of Sciences/Beijing National Laboratory for Condensed Matter Physics, Beijing 100190, China}
\author{Hongsheng Chen}
\affiliation{State Key Laboratory of Modern Optical Instrumentation, College of Information Science and Electronic Engineering, Zhejiang University, Hangzhou 310027, China}
\author{Zhong Wang}
\affiliation{Institute for Advanced Study, Tsinghua University, Beijing 100084, China}
\affiliation{Collaborative Innovation Center of Quantum Matter, Beijing 100871, China}
\author{Ling Lu}  \email{linglu@iphy.ac.cn\\$^{\dagger}$The first two authors contributed equally to this work.}
\affiliation{Institute of Physics, Chinese Academy of Sciences/Beijing National Laboratory for Condensed Matter Physics, Beijing 100190, China}

\maketitle

\textbf{
Three-dimensional~(3D) topological nodal points~\cite{Armitage2017Weyl,wan2011,xu2015discovery,lv2015experimental,lu2015,soluyanov2015type}, such as Weyl and Dirac nodes have attracted 
wide-spread interest across multiple disciplines and diverse material systems. Unlike nodal points that contain little structural variations, nodal lines~\cite{burkov2011topological,lu2013weyl,fang2016topological} can have numerous topological configurations in the momentum space, forming nodal rings~\cite{kim2015dirac,yu2015topological,fang2015topological,kobayashi2017crossing}, nodal chains~\cite{bzduvsek2016nodal,kawakami2016symmetry,yu2017nodal,wang2017HourglassDiracChain,feng2017topological} and potentially nodal links~\cite{chen2017topological,yan2017nodal,chang2017weyl} and nodal knots~\cite{ezawa2017topological,bi2017nodal}.
However, nodal lines have much less development for the lack of ideal material platforms~\cite{bian2016topological,schoop2016dirac,chen2017dirac}. In condensed matter for example, nodal lines are often fragile to spin-orbit-coupling, locating off the Fermi level, coexisting with energy-degenerate trivial bands and dispersing strongly in energy of the line degeneracy. Here, overcoming all above difficulties, we theoretically predict and experimentally observe nodal chains in a metallic-mesh photonic crystal having frequency-isolated linear band-touching rings chained across the entire Brillouin zone~(BZ).
These nodal chains are protected by mirror symmetries and have a frequency variation less than 1\%.
We used angle-resolved transmission~(ART) to probe the projected bulk dispersions and performed Fourier-transformed field scan~(FTFS) to map out the surface dispersions, which is a quadratic touching between two drumhead surface bands.
Our results established an ideal nodal-line material for further studies of topological line-degeneracies with nontrivial connectivities, as well as the consequent wave dynamics richer than 2D Dirac and 3D Weyl materials.
}

\begin{figure}[t]
\includegraphics[width=0.4 \textwidth]{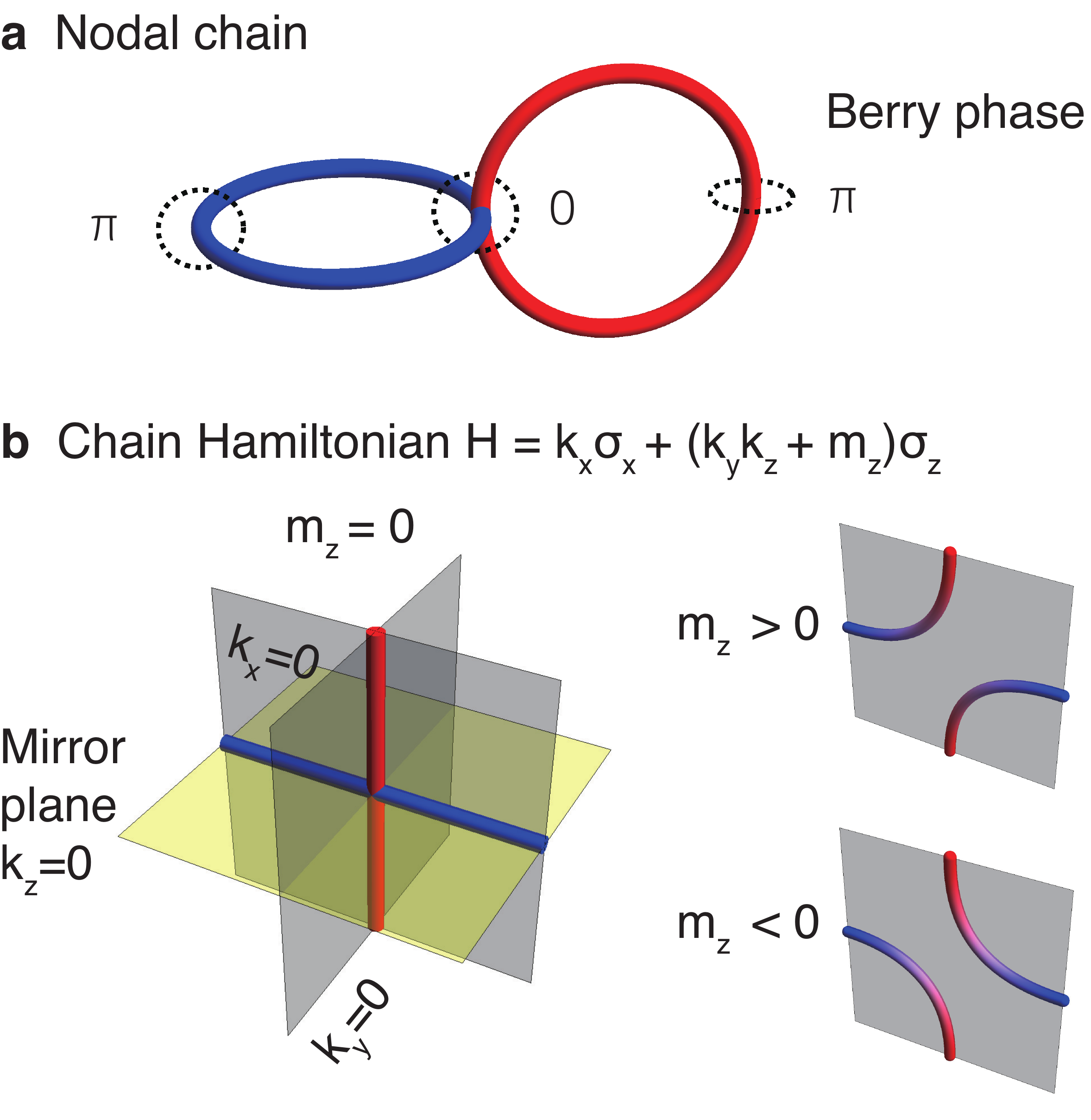}
\caption{Nodal-chain Hamiltonian and stability.
\textbf{a}, Illustration of the simplest chain structure between two rings. The Berry phase around the chain point is 0, in contrast to the $\pi$ Berry phase of nodal lines.
\textbf{b}, The chain point is the crossing between two nodal lines defined between three zero planes. The third plane in yellow represents the mirror plane protecting the chain point. When the mirror symmetry is broken by the mass term $m_z$, the chain point splits.}
\label{fig1}
\end{figure}

\paragraph{Chain Hamiltonian}
Nodal line is the extrusion of a Dirac cone, arguably the most intriguing 2D band structure, into 3D momentum space. They share the same local Hamiltonian $H(\mathbf{k})=k_x\sigma_x + k_y\sigma_z$ that can be protected by the $\mathcal{PT}$ symmetry forbidding the mass term of $\sigma_y$ in the whole BZ, where $\mathcal{P}$ is parity inversion and $\mathcal{T}$ is time-reversal symmetry.
A single nodal line usually form a closed ring, due to the periodicity of the BZ. Surprisingly, it was recently proposed that nodal rings can be chained together as shown in Fig. \ref{fig1}a.
Other than $\mathcal{PT}$, the critical chain point requires an extra symmetry to be stabilized. For example, such symmetry can be glide or mirror planes. This is clear in the chain Hamiltonian $H(\mathbf{k})=k_x\sigma_x + (k_yk_z+m_z)\sigma_z$ that we propose here and plot in Fig. \ref{fig1}b.
When $m_z=0$, it defines two nodal lines crossing at the origin. The red nodal line locates at the intersection between the planes of $k_x=0$ and $k_y=0$ and the blue nodal line locates at the intersection between the planes of $k_x=0$ and $k_z=0$.
The vanishing mass $m_z=0$ is guaranteed by the mirror~(glide) symmetry $M_z=\sigma_x$ that flips $z$ coordinates, shown in Fig. \ref{fig1}b as a yellow plane. When the mirror symmetry is broken, the chain point is lifted and the nodal lines become hyperbolic.

\begin{figure*}[t]
\includegraphics[width=0.8\textwidth]{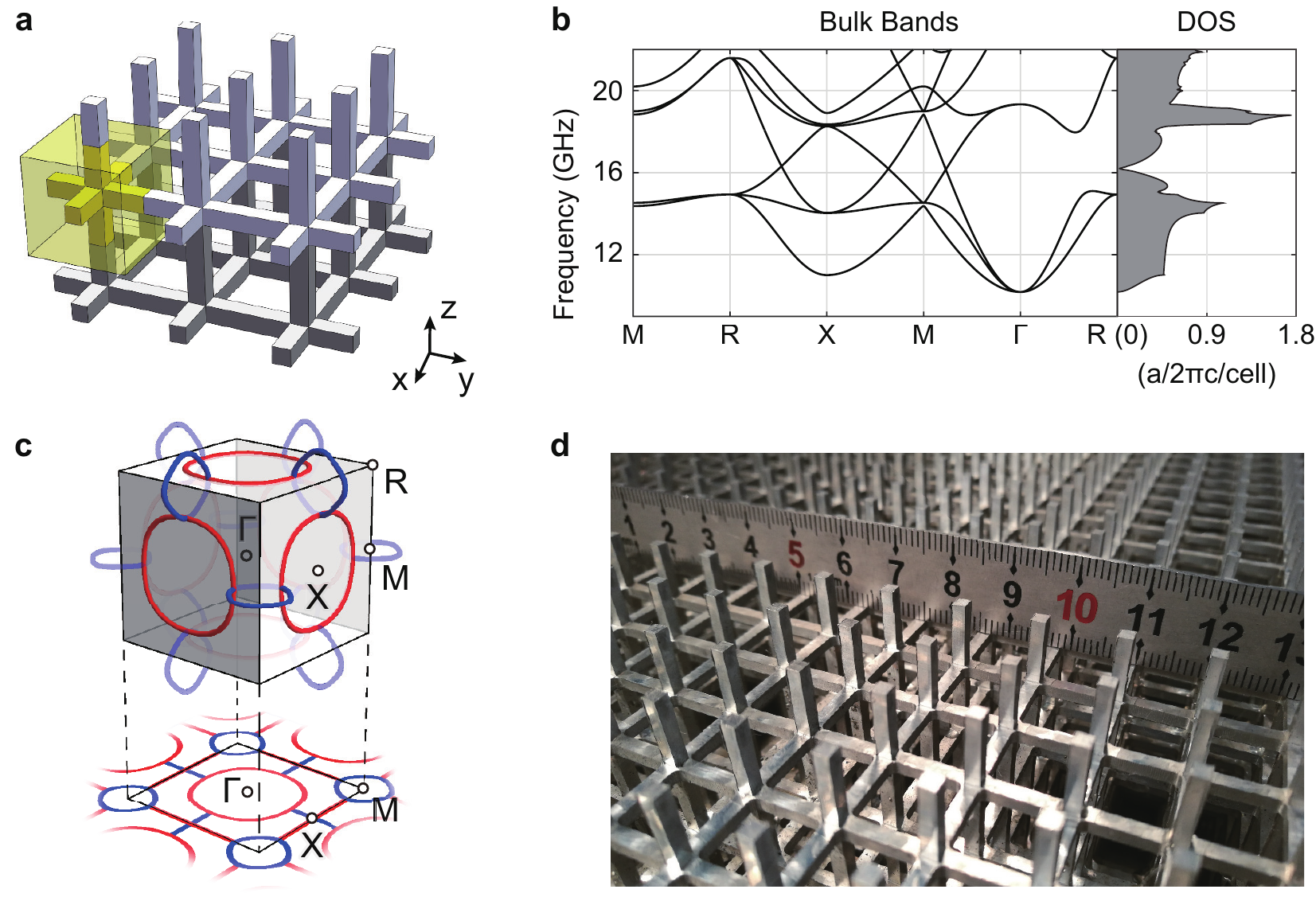}
\caption{Nodal-chain photonic crystal.
\textbf{a}, The illustration of the metallic mesh 3D photonic crystal. The yellow cube denotes a symmetric simple-cubic unit cell.
\textbf{b}, The bulk band structure and DOS. 
The nodal-chain frequency is $\sim$16.2GHz.
\textbf{c}, The structure of nodal chains in the BZ. The three blue rings are chained with the three red rings.
\textbf{d}, The top surface of the sample made of aluminum alloy. The bottom surface is a flat mesh. The lattice constant is 11.6mm and the rod width is 2mm.
}
\label{fig2}
\end{figure*}

\begin{figure*}[t]
\includegraphics[width=0.9\textwidth]{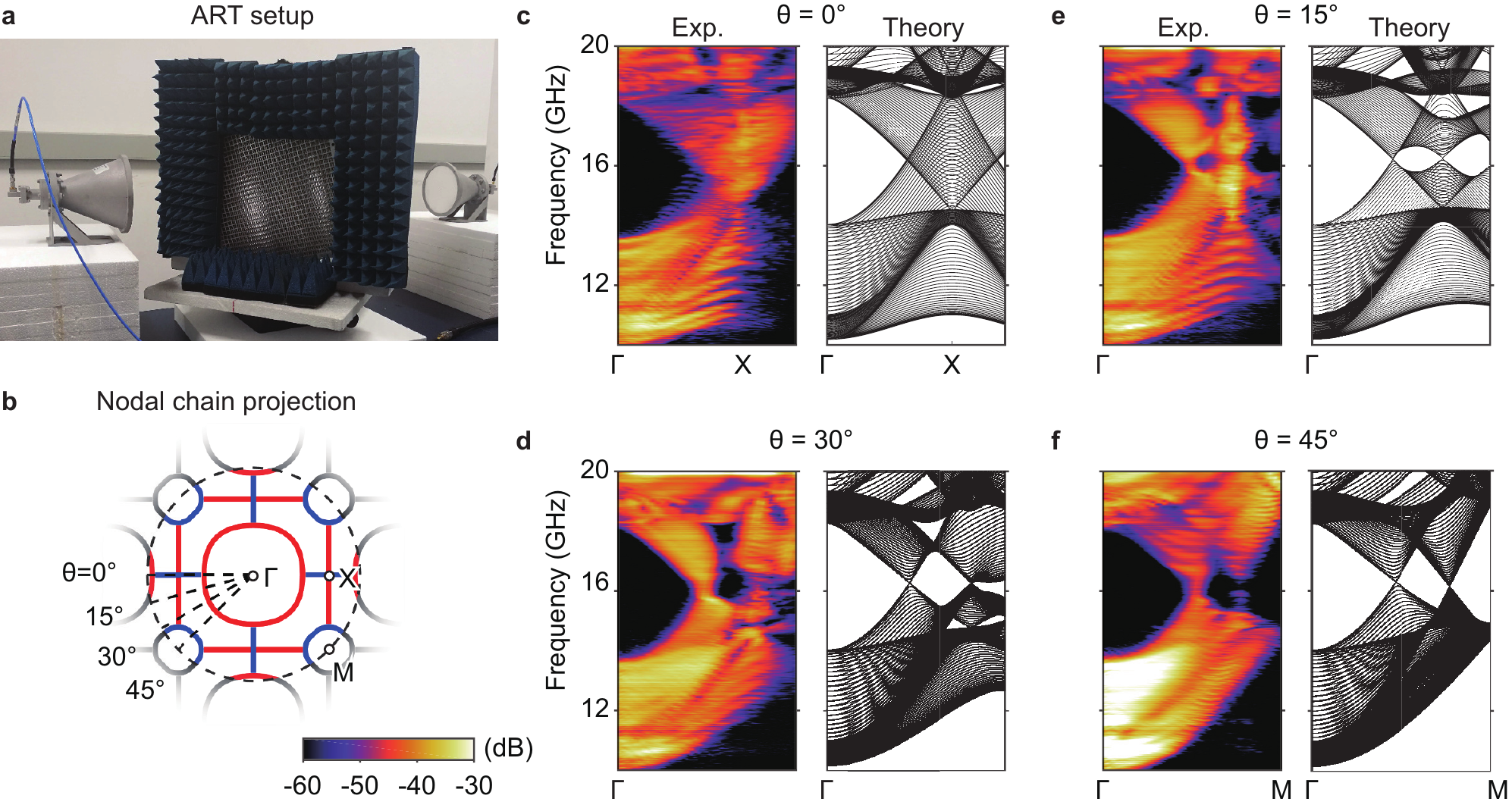}
\caption{ART measurement of nodal-chain bulk states.
\textbf{a}, The experiment setup. 
\textbf{b}, Projective view of the nodal chain in (001) direction. The dashed circle is the light cone in prism at 16.2 GHz, which is the maximum momentum of the incident photon.
Four dashed radii denote the four scan directions.
\textbf{c},\textbf{d},\textbf{e},\textbf{f}
Bulk experimental data compared with theoretical calculations for differnet $\theta$ values. The agreement is great.}
\label{fig3}
\end{figure*}

\paragraph{Nodal-chain photonic crystal}
We designed a metallic-mesh 3D photonic crystal having frequency-isolated nodal chains. The structure is shown in Fig. \ref{fig2}a. The yellow box denotes the unit cell that belongs to the space group $Pn\overline{3}n$, no. 221. Each unit cell consists of three square rods, along x, y, and z directions, intersecting at the center of the unit cell. Numerically, we treat the metallic surfaces as perfect electric conductors~(PEC), which is a good approximation for noble metals at microwave frequencies.

Interesting enough, we note that similar metallic mesh structure has also been used for low-frequency plasmons~\cite{pendry1996extremely}, particle accelerators~\cite{shapiro20053d}, novel metamaterials~\cite{chen2016metamaterials} as well as invisible materials\cite{ye2016invisible}.  

Shown in Fig. \ref{fig2}b, the third and the fourth bands cross linearly, forming line degeneracies of nodal chains.
The chain structure is plotted in Fig. \ref{fig2}c, where a total of six nodal rings are chained in the BZ.
There are three red rings centered at the \emph{X} points on each face and three blue rings centered at the \emph{M} points on each edge. The same colored rings are related by the (111) rotation symmetry.
Each blue ring is chained with the neighboring red rings perpendicular to it. All chain points lie in and are protected by the \{001\} mirror planes.

An important feature of this design is the low dispersion of nodal-line frequencies, which is below 1\% for the variation~($\Delta\omega/\omega_{middle}$) in the entire BZ. This is supported by the density of states~(DOS) calculations on the right of Fig. \ref{fig2}b. There is a clear dip in the DOS increasing linearly away from the nodal-chain frequency.

\paragraph{Sample fabrication}
In experiment, we adopted aluminum as the material of choice for its high conductivity, light weight and low cost. The sample is shown in Fig. \ref{fig2}a with a lattice constant of 11.6mm and a rod width of 2mm.
The resulting nodal-chain frequencies are very close to 16.2 GHz, as shown in Fig. \ref{fig2}b.
The sample is stacked by 9 identical layers and each layer has 30$\times$30 unit cells. Every layer was milled and drilled from a plain aluminum plate. For assembling, handling and alignment, we reserved frames around each layer. The final size for one layer is 37cm$\times$37cm$\times$10.44cm.

\paragraph{Bulk state measurement}
We performed ART to measure the nodal-chain bulk states.
As shown in Fig. \ref{fig3}a, we used a similar setup as in Ref. \cite{lu2015}. 
In this method, we detect the frequency-resolved transmission as a function of incident angle that converts to momentum.
As a result, ART measures the projected bulk states along the normal direction of the sample surface, which is the z axis in our experiment. Figure \ref{fig3}b shows the (001) projection of nodal chains from numerical calculations ~(consistent with Fig. \ref{fig2}c).
$\theta$ was sampled from 0 degree to 45 degrees, with a step size of 15 degrees. 
A vector network analyzer was used to collect data. A pair of prisms, with refractive index of 4, was applied to enlarge the scanning range in the reciprocal space~(same as those in Ref. \cite{lu2015}). For comparison, we presented the results without the prisms in Supplementary Information.

The comparisons between experimental results and projected band structures are shown in Fig. \ref{fig3}c-f for four different $\theta$ angles. With an over 25dB attenuation in bandgaps, results are clear and in good agreement with theory. The transmission data for each polarization channel are also presented in the Supplementary Information.

\begin{figure*}[t]
\includegraphics[width=\textwidth]{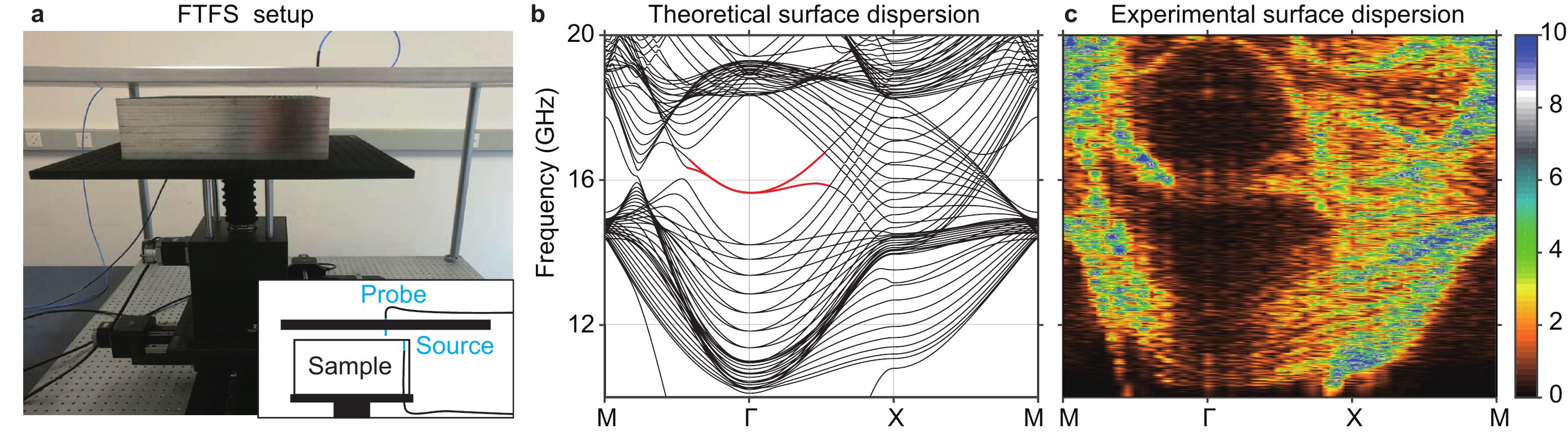}
\caption{FTFS measurement of the drumhead surface states. We measured the bottom surface of flat mesh.
\textbf{a}, Picture and schematic illustration of the 2D profiler.
\textbf{b}, Theoretical calculation of the surface band structure. The supercell consists of 11 unit cells, with an 8mm air gap to the upper PEC boundary. Plotted in red, the two drumhead surface dispersions are degenerate at the $\Gamma$ point of the surface BZ.
\textbf{c}, The experimental result plotted in the linear scale. Both bulk modes and surface modes are clearly seen. The surface modes appear at $\sim$16GHz that matches the theoretical prediction.}
\label{fig4}
\end{figure*}

\paragraph{Surface state measurement}
We performed FTFS to measure the quadratic-touching of drumhead surface states of our nodal-chain sample.
A nodal-line material is known to support drumhead surface state, that is a sheet of surface dispersion enclosed by the projected nodal-line bulk states in the surface BZ.
We form a surface by placing an aluminum plate 8mm above the sample (001) surface, in order to isolate the free space photon modes while leaving space for the near-field scanning. The setup is shown in Fig. \ref{fig4}a.
Due to the $C_4$ symmetry on our sample surface, we numerically discovered two drumhead surface dispersions forming a quadratic degeneracy at the center of the surface BZ, shown in Fig. \ref{fig4}b. The surface state is highlighted in red lines.

During the field scan, the broadband source was fixed inside the sample to excite both the surface and the bulk states. The probe was placed inside the hole at the center of the aluminum board. The detailed parameters of the profiler are presented in Supplementary Information. During the measurement, the sample moved in-plane on the guiding rails and the probe scanned across the sample surface point by point. 
Both the amplitude and the phase of the local fields were recorded and frequency-resolved by the network analyzer.

We then Fourier-transformed the field data from real space to reciprocal space and summed up the amplitudes in all equivalent Bloch momenta to the first BZ. The resulting intensity map can be directly compared to the dispersion calculations.
Here we plotted the intensity along a 1D path ($M-\Gamma-X-M$), as shown in Fig. \ref{fig4}c. Surface modes exist at $\sim$16GHz and are degenerate at the $\Gamma$ point.

\paragraph{Conclusion}
We theoretically identified and experimentally verified a nearly-ideal nodal-chain photonic crystal, which will motivate further experiments of such materials in electron, and phonon systems, as well as photonic realizations towards optical frequencies~\cite{lu2016symmetry,chen2016photonic,slobozhanyuk2017three,noh2017experimental}.
The search for nodal-link and nodal-knot materials will also be an exciting direction.

\paragraph{Acknowledgments}
We thank Chen Fang and Hongming Weng for discussions.
The authors were supported by the Ministry of Science and Technology of China under Grant No. 2017YFA0303800 and 2016YFA0302400 (Q.Y, L.L.), NSFC under Grant No. 11674189 (Z.Y., Z.W.), 
the National Natural Science Foundation of China under Grants No. 61625502~(H.C.), No. 61574127~(H.C.), the Top-Notch Young Talents Program~(H.C.) and National Thousand-Young Talents
Program~(L.L.) of China.

\bibliography{reference}

\end{document}